\documentclass[journal = nanolett, manuscript = article]{achemso}
\usepackage{setspace}
\setkeys{acs}{articletitle = true, etalmode=truncate, maxauthors = 10}
\usepackage{graphicx}
\usepackage{amsmath,amssymb,amsfonts,bm}
\usepackage{mathtools}
\usepackage[colorlinks,citecolor=blue,urlcolor=black]{hyperref}
\usepackage{braket}
\author{Sandip Aryal}
\affiliation{Department of Chemistry, Wayne State University, Detroit, Michigan 48202, USA}
\author{Joseph Frimpong}
\affiliation{Department of Chemistry, Wayne State University, Detroit, Michigan 48202, USA}
\author{Zhen-Fei Liu}
\email{zfliu@wayne.edu}
\affiliation{Department of Chemistry, Wayne State University, Detroit, Michigan 48202, USA}
\title{Comparative Study of Covalent and van der Waals CdS Quantum Dot Assemblies from Many-Body Perturbation Theory}

\begin{document}

\begin{abstract}
Quantum dot (QD) assemblies are nanostructured networks made from aggregates of QDs and feature improved charge and energy transfer efficiencies compared to discrete QDs. Using first-principles many-body perturbation theory, we systematically compare the electronic and optical properties of two types of CdS QD assemblies that have been experimentally investigated: QD gels, where individual QDs are covalently connected via di- or poly-sulfide bonds, and QD nanocrystals, where individual QDs are bound via van der Waals interactions. Our work illustrates how the electronic, excitonic, and optical properties evolve when discrete QDs are assembled into 1D, 2D, and 3D gels and nanocrystals, as well as how the one-body and many-body interactions in these systems impact the trends as the dimensionality of the assembly increases. Furthermore, our work reveals the crucial role of the covalent di- or poly-sulfide bonds in the localization of the excitons, which highlights the difference between QD gels and QD nanocrystals.

\end{abstract} 

%\section{Introduction}
% The importance of QD, and the need of assemblies
Quantum dots (QDs) are a class of spatially confined materials, usually a few nanometers in size. The quantum confinement leads to size-dependent electronic properties \cite{B84, EEO85,NERB96,CMCM06,RKK07,MLSM09} that are more similar to atoms (hence the term ``artificial atom'') than bulk semiconductors \cite{SB90a,TZ11}. QDs feature high energy conversion efficiency \cite{LMS12,LHMJ20} and the tunability of their photophysical and photochemical properties makes them suitable for a wide range of applications, especially in solar cells \cite{SLB12, KLYJ20, PTKA21}, light-emitting diodes \cite{AB07, RBBM18, NCC19}, transistors \cite{TM05,ZGC98}, and photocatalysis \cite{BTCA07,MCSW20,JW20}, just to name a few. However, QDs often suffer from photodegradation and photocorrosion \cite{JSLY13,AZZT18,MLLK19}, luminescence quenching \cite{BHN07,ZRPK12}, as well as limited carrier and energy transfer efficiencies \cite{TKHJ11, WLGE12}. Efforts have been made to address these issues via functionalization or passivation of QD surfaces using ligands \cite{TKHJ11, BJPW12, ZSZB14, CSLS16, BLHT16a} and molecular catalysts \cite{KLSW18,MCSW20}.

% Why do we need gels / nanocrystals. Missing of understanding of structure-property relationship (different dimensions, etc).
QD assemblies \cite{MKB95, CVH98, BWSG09, NBLL10, CJDW15, SEE13}, macroscopic architectures made from aggregates of QDs, present an attractive solution to the limitations mentioned above \cite{OEKT15, BET16, WYSK16, YAYF20}. Without the often bulky organic ligands, QD assemblies provide an excellent path to connect individual QDs to an ``all inorganic'' network that is a suitable platform for large devices, while improving charge and energy transfer efficiencies compared to discrete QDs \cite{MLDJ02, CRVC11, WLGE12, WYSK16}. Different types of QD assemblies based on II-VI materials have been investigated, including those in which individual QDs are coupled by chemical bonds between surface atoms \cite{GLLB01, AB07b, ZWLH17, HJLK19, HGSN20, GOKA22, HSBL21, SGHA21, DHMV20}, and those in which individual QDs interact via van der Waals forces \cite{MKB95, CVH98, BET16, MGG19}. The former features interconnected pore structures between QDs \cite{AMB05,MAB05,BK10,ZWLH17}, exhibits fractal dimensionality \cite{GLLB01}, and is termed ``QD gel'' in this work. The latter features periodic arrays of QDs arranged in a superlattice \cite{BDFP15, KMCH15, CBA22} and is termed ``QD nanocrystal (NC)'' in this work.

% Prior computational work on QD, and no QD gels.
Given the improved characteristics of QD assemblies compared to discrete QDs, it is imperative to understand how the structural differences between discrete QDs and different types of QD assemblies lead to distinct properties. To this end, first-principles calculations provide a powerful means to reveal the microscopic structure-property relationship, complementary to various experimental techniques. Most prior computational studies focused on discrete QDs, such as those on the effect of doping \cite{ZFB03, GL16, BGVG17}, the role of passivation \cite{HLC05, FJS05}, and the size- and shape-dependent electronic \cite{WZ94, JSS00, LW03, WL04, CRVC11} and optical \cite{WZ94-PRL,MCFZ97,FZ97,FP02,FJS07,FTSJ11,EC13} properties. Additionally, prior studies on the QD NCs \cite{VGZ15,MGG19} illustrated the roles played by the ordered arrays and superlattices. However, we have not found similar studies of covalent QD gels. Additionally, a complete account of the differences between the two types of QD assemblies in different dimensions, and more importantly, a precise and microscopic understanding of the difference in the quantum confinement between discrete QDs and QD assemblies, are missing. These knowledge gaps hinder future development of QD assemblies as energy materials.

% In this work...
In this work, we leverage first-principles calculations to illustrate the structure-property relationships in band gaps and optical properties for a series of QD assemblies, including QD gels and QD NCs formed in 1D, 2D, and 3D, and compare them with a discrete QD. All the structures studied in this work are constructed from a prototypical spherical CdS QD with a 1.6 nm diameter, which is found to be stable in experiments and is one of the stoichiometric sizes \cite{KNN05} that allow a charge-orbital balance\cite{VZSN12} when passivated. Due to known issues in the calculation of band gaps and excitonic properties associated with most density functionals, \cite{ORR02} we employ the first-principles $GW$-BSE formalism \cite{H65, HL86, RL00} ($G$: Green's function; $W$: screened Coulomb interaction; BSE: Bethe-Salpeter equation) within the framework of many-body perturbation theory, which is state-of-the-art for computing quasiparticle and optical properties \cite{GDR19,BDJL20}. $GW$-BSE has been successfully applied to discrete QDs in the literature \cite{OCL97,W05,PTC06,PTC08,FITR09,WS14,ZCKR15,KBBL18}. Here, we aim to unveil the differences between a discrete QD, QD gels, and QD NCs, and discuss the interplay between quantum confinement and many-body effects, as well as the unique roles of the di- and poly-sulfide linkers in QD gels.

%\section{Computational Details} 
% Fig1: structures (four panel: dot, 1D 2S, 1D 4S, 1D nanocrystal)
We start with creating a spherical CdS QD of 1.6 nm diameter from the bulk wurtzite CdS. Direct calculations of this QD lead to mid-gap states that stem from the dangling bonds of the surface atoms, which is different from the experimental condition where the QD surface is often passivated by ligands. In our work, we adopt the passivation scheme following Refs. \citenum{WL04, HLC05}, where the QD surface atoms are passivated with pseudo-hydrogen atoms with $(8-m)/4$ electrons with $m$ being the number of valence electrons of a surface atom. Within this scheme, surface Cd (S) atoms are passivated with pseudo-hydrogen atoms with 1.5 (0.5) electrons to remove the dangling bonds and the resulting mid-gap states while maintaining charge neutrality for the QD.

The discrete QD structure with pseudo-hydrogen atoms is relaxed using the local density approximation (LDA) \cite{KS65,PZ81} within the framework of density functional theory (DFT), as implemented in the Quantum ESPRESSO package.\cite{GBBB20} We use the projector augmented wave (PAW) method in the geometry relaxations and the optimized norm-conserving Vanderbilt pseudopotentials (ONCV) \cite{H13, SG15} in the subsequent single-point electronic structure calculations. The relaxation uses a kinetic energy cutoff of 50 Ry and a simulation cell of 35 \AA~along each direction until all residual forces are below 0.05 eV/\AA.

To model QD gels, we connect neighboring QDs with di- or poly-sulfide covalent bonds to form periodic structures, in line with the experimental gelation procedure \cite{AB07,AMB05}. We consider two microscopic models: (1) ``2S'' gels, where a disulfide bond connects two neighboring QDs, with one sulfur atom embedded on the surface of one QD and the other sulfur atom attached to a surface Cd atom of the other QD; and (2) ``4S'' gels, where a tetrasulfide bond connects two neighboring QDs, with one sulfur atom terminus embedded on the surface of one QD and the other terminal of the tetrasulfide chain attached to a surface Cd atom of the other QD. In both cases, we remove the passivating pseudo-hydrogen atoms from the surface Cd or S atoms that are directly connected to the di- or tetrasulfide bonds. We consider gels formed in 1D, 2D, and 3D, respectively, in simple cubic lattices, as limits of the realistic gels in fractal dimensions \cite{YLB09}. This simplification in the modeling allows us to unambiguously examine the effect of the dimensionality in modulating electronic and optical properties. We perform variable-cell relaxations along the periodic direction(s), to fully relax the local binding geometry of the di- or tetrasulfide bonds, and use a size of 35 \AA~for the simulation cell in non-periodic direction(s). 

To model QD NCs in different dimensions, we place neighboring QDs in close contact (within a few \AA's) along the periodic directions and use a size of 35 \AA~for the simulation cell in non-periodic directions to form simple cubic superlattices, before we start the variable-cell relaxations. Here, no covalent bonds exist between QDs, and we do not remove any pseudo-hydrogen atoms from the surface. In the 1D NC, the relaxed inter-dot distance as measured between two pseudo-hydrogen atoms attached to neighboring QD surfaces is about 3.7 \AA, which translates to about 6.5 \AA~between surface Cd/S atoms on neighboring QDs. As a comparison, the relaxed inter-dot distance (as measured between surface Cd/S atoms on neighboring QDs) is about 3.8 \AA~in the 1D 2S gel and 4.2 \AA~in the 1D 4S gel. 

\begin{figure}[htb!]
\centering
\includegraphics[width=4in]{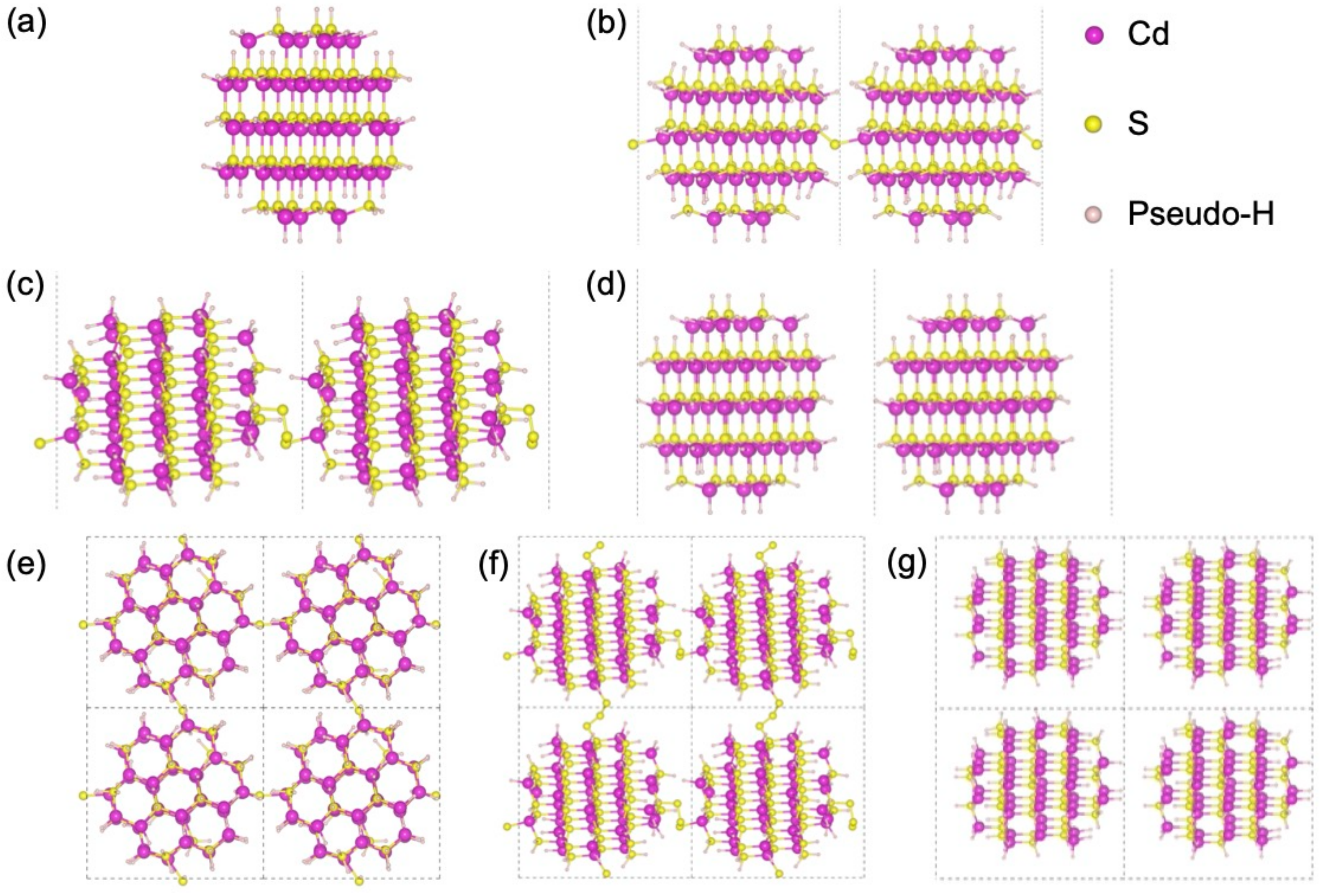}
\caption{(a) A discrete CdS QD of 1.6 nm diameter, with pseudo-hydrogens on the surface as passivation. (b) 1D 2S QD gel. (c) 1D 4S QD gel. (d) 1D QD NC. (e) 2D 2S QD gel. (f) 2D 4S QD gel. (g) 2D QD NC. For the gel (NC) structures, two neighboring unit cells along each periodic direction are displayed to demonstrate the presence (absence) of di- or tetrasulfide bonds between the QDs. In all panels, the dashed lines are the boundaries of the simulation cells.}
\label{fig:stru}
\end{figure}

The optimized structures of the discrete QD, the 2S gels, 4S gels, and NCs in 1D and 2D are shown in Figure \ref{fig:stru}. For the gel (NC) structures, two neighboring unit cells along each periodic direction are displayed to demonstrate the presence (absence) of the di- or tetrasulfide bonds between the QDs. The relaxed lattice parameters for all systems studied in this work are summarized in Table S1. 

\begin{figure}[htb]
\centering
\includegraphics[width=4in]{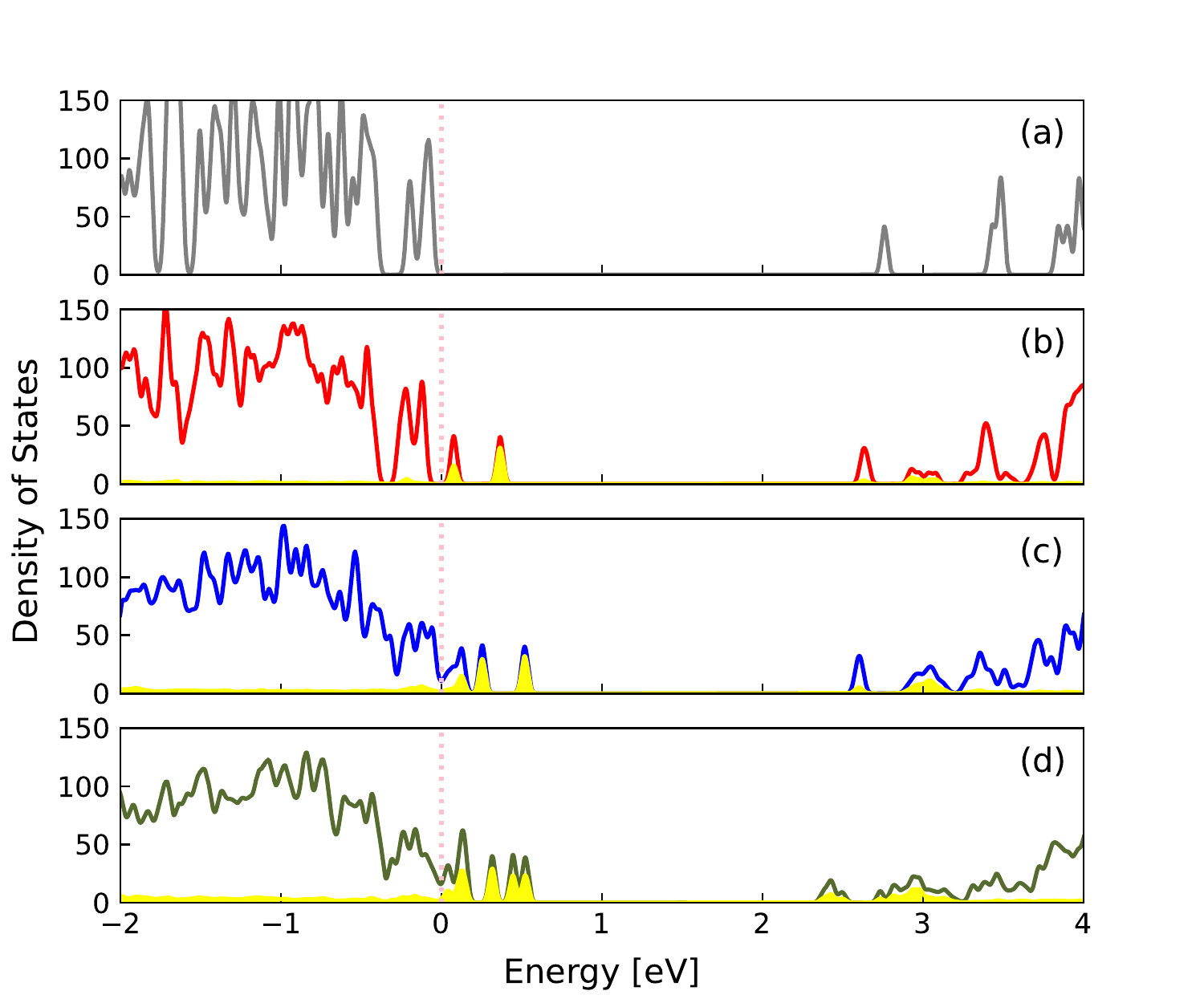}
\caption{DOS calculated from DFT-LDA, for (a) the discrete QD, (b) 1D 2S QD gel, (c) 2D 2S QD gel, and (d) 3D 2S QD gel. Yellow shaded areas highlight the projected DOS onto the two linker sulfur atoms. All panels are aligned at the energy where the orbital is most similar to the VBM of the discrete QD, as indicated by the pink dashed line.}
\label{fig:dos}
\end{figure}

Figure \ref{fig:dos} shows the density of states (DOS) calculated from DFT-LDA, comparing the discrete QD, 1D, 2D, and 3D 2S QD gel. Similar results are shown in Figure S1 for the 4S QD gels and in Figure S2 for the QD NCs in different dimensions. Solid lines show the total DOS and yellow shaded areas highlight the projected DOS onto the two sulfur atoms that link neighboring QDs in the gel. To facilitate a comparison, we have aligned all panels in Figure \ref{fig:dos} at the energy where the orbital is most similar to the valance band maximum (VBM) of the discrete QD, as indicated by the pink dashed line. Compared to the discrete QD, covalently bound gels feature additional ``mid-gap'' states above the VBM of the discrete QD, which are localized on the linker sulfur atoms. The conduction band minimum (CBM) of the gels, however, is still largely localized on the QD rather than on the linker. The appearance of these ``mid-gap'' states effectively reduces the band gaps, as we show in Table \ref{tab:gaps} below. On the contrary, QD NCs, bound via van der Waals interactions, have LDA band-gap values similar to that of the discrete QD (the \emph{many-body} gaps, or the physical gaps, however, do differ from that of the discrete QD, see below), as shown in Table \ref{tab:gaps} and Figure S2. This distinction between gels and NCs underlines the effect of the covalent sulfur linkers in modulating the electronic structure. We note that for all periodic systems, the DFT-LDA band structures exhibit a weak dispersion (around 0.15 eV), with a direct band gap found at the edge of the Brillouin zone, as we show in Figure S3. 

% dot vs 1D vs 2D vs 3D (2S) DOS - Fig 2

For a quantitatively accurate description of the electronic and optical properties, we turn to first-principles $GW$-BSE calculations as implemented in the BerkeleyGW package\cite{DSSJ12}. The LDA electronic structure discussed above is used as the starting point, and we compute the self-energies perturbatively, i.e., $G_0W_0$@LDA. The dielectric cutoff and the number of bands included in the calculation of the non-interacting polarizability are determined from convergence studies (see Table S2 and Figure S4), and we list the computational parameters used for each system in Table S1. To remove the spurious Coulomb interactions between different images along the non-periodic directions, we apply the box truncation for the discrete QD, the wire truncation for 1D assemblies, and the slab truncation for 2D assemblies. In the self-energy calculations, we treat the frequency dependence using the Hybertson-Louie generalized plasmon pole model \cite{HL86}, and apply the static reminder approximation \cite{DSJC13} to speed up the convergence. In the BSE calculations, we include 20 valence bands and 20 conduction bands in the active space to construct the BSE Hamiltonian, which is found to converge the absorption spectrum (see Figure S5).

\begin{table}
\caption{Summary of the electronic and optical properties for all systems studied in this work. $E_{\rm g}^{\rm LDA}$ ($E_{\rm g}^{GW}$) is the transport gap calculated from DFT-LDA ($GW$). For any QD assembly system $A$, $\Delta_{\rm g}^{\rm MF}(A)=E_{\rm g}^{\rm LDA}\mbox{(QD)}-E_{\rm g}^{\rm LDA}(A)$, and $\Delta_{\rm g}^{\rm MB}(A)=E_{\rm g}^{GW}\mbox{(QD)}-E_{\rm g}^{GW}(A)-\Delta_{\rm g}^{\rm MF}(A)$. $E_1$ ($E_2$) is the first (second) peak in the BSE optical spectra, shown as the solid curves in Figure \ref{fig:bse}. $E_{\rm b}$ is the exciton binding energy, defined as $E_{\rm g}^{GW}-E_1$. All energies are in eV.}
\begin{tabular}{c|ccccccc}
\hline
\hline
System & $E_{\rm g}^{\rm LDA}$ & $E_{\rm g}^{GW}$ & $\Delta_{\rm g}^{\rm MF}$ & $\Delta_{\rm g}^{\rm MB}$ & $E_{\rm 1}$ & $E_{\rm 2}$ & $E_{\rm b}$\\
\hline
\hline
QD & 2.83 & 5.73 &  $-$ & $-$ & 4.02 & $-$ & 1.71\\
\hline
1D 2S gel & 2.25 & 4.94 & 0.58 & 0.21 & 3.08 & 3.85 & 1.86 \\
2D 2S gel & 2.05 & 4.58 & 0.78 & 0.37 & 2.88 & 3.74 & 1.70\\
3D 2S gel & 1.83 & 3.82 & 1.00 & 0.91 & 2.64 & 3.67& 1.18\\
\hline
1D 4S gel & 2.34 & 5.02 & 0.49 & 0.22 & 3.24 & 3.93 & 1.78\\
2D 4S gel & 2.02 & 4.59 & 0.81 & 0.33 & 3.19 & 3.94 & 1.40\\
3D 4S gel & 1.89 & 4.03 & 0.94 & 0.76 & 3.27 & 4.10 & 0.76\\
\hline
1D NC & 2.83 & 5.68 & 0.00 & 0.05 & 4.02 & $-$ & 1.66\\
2D NC & 2.82 & 5.62 & 0.01 & 0.10 & 4.01 & $-$ & 1.61\\
3D NC & 2.82 & 5.39 & 0.01 & 0.33 & 4.14 & $-$ & 1.25\\
\hline
\hline
\end{tabular}
\label{tab:gaps}
\end{table}

Table \ref{tab:gaps} compares the transport (fundamental) band gaps calculated from DFT-LDA ($E_{\rm g}^{\rm LDA}$) and $GW$ ($E_{\rm g}^{GW}$), defined as the difference between the VBM and the CBM energies. These results reveal the mean-field and many-body effects of the di- and tetrasulfide linkers in modulating the band gaps of QD assemblies compared to the discrete QD. As an example, comparing the 1D 2S gel with the discrete QD, the DFT-LDA gap decreases by 0.58 eV, while the $GW$ gap decreases by 0.79 eV. The former is denoted by $\Delta_{\rm g}^{\rm MF}$ in Table \ref{tab:gaps} and is a result of the covalent bond and the formation of the ``mid-gap'' states, hence can be captured by mean-field theories such as LDA. On the other hand, the \emph{additional} 0.21 eV change in the $GW$ gap, denoted by $\Delta_{\rm g}^{\rm MB}$ in Table \ref{tab:gaps}, is a genuine many-body effect: neighboring QDs in the periodic 1D gel act as a dielectric environment, providing screening of the Coulomb interaction within one QD and reducing the band gap. This is the same physical effect that explains the gap difference between a molecular crystal and a single molecule \cite{RSJB13}, and can only be correctly captured by beyond-mean-field techniques such as the $GW$ method used here.

Table \ref{tab:gaps} unveils interesting trends across all the systems we study. First, for both the 2S and 4S gels, when the dimensionality increases, the $\Delta_{\rm g}^{\rm MF}$ increases. This is due to the reduced quantum confinement and the additional linker sulfur atoms present in the system for higher dimensions, which introduce additional ``mid-gap'' states, as shown in Figure \ref{fig:dos}. Second, the $\Delta_{\rm g}^{\rm MB}$ also increases as dimension, due to enhanced dielectric screening as more neighboring QDs are present when the dimensionality increases. Furthermore, the values of both $\Delta_{\rm g}^{\rm MF}$ and $\Delta_{\rm g}^{\rm MB}$ are similar (generally within 0.1 eV) for the 2S and 4S gels of the same dimension, because they stem from the same physical effect and the inter-dot distance is similar for the 2S and 4S gels. Third, the $\Delta_{\rm g}^{\rm MF}$ is uniformly zero for the NCs in all dimensions, due to the absence of covalent bonds connecting neighboring QDs. Fourth, the $\Delta_{\rm g}^{\rm MB}$ for the NCs increases as dimension, consistent with the trends observed in the covalent gels. But the values here are much smaller than the gels, due to the larger inter-dot distance in the NCs, resulting in weaker dielectric screening.

\begin{figure*}[htb!]
\centering
\includegraphics[width=6.3in]{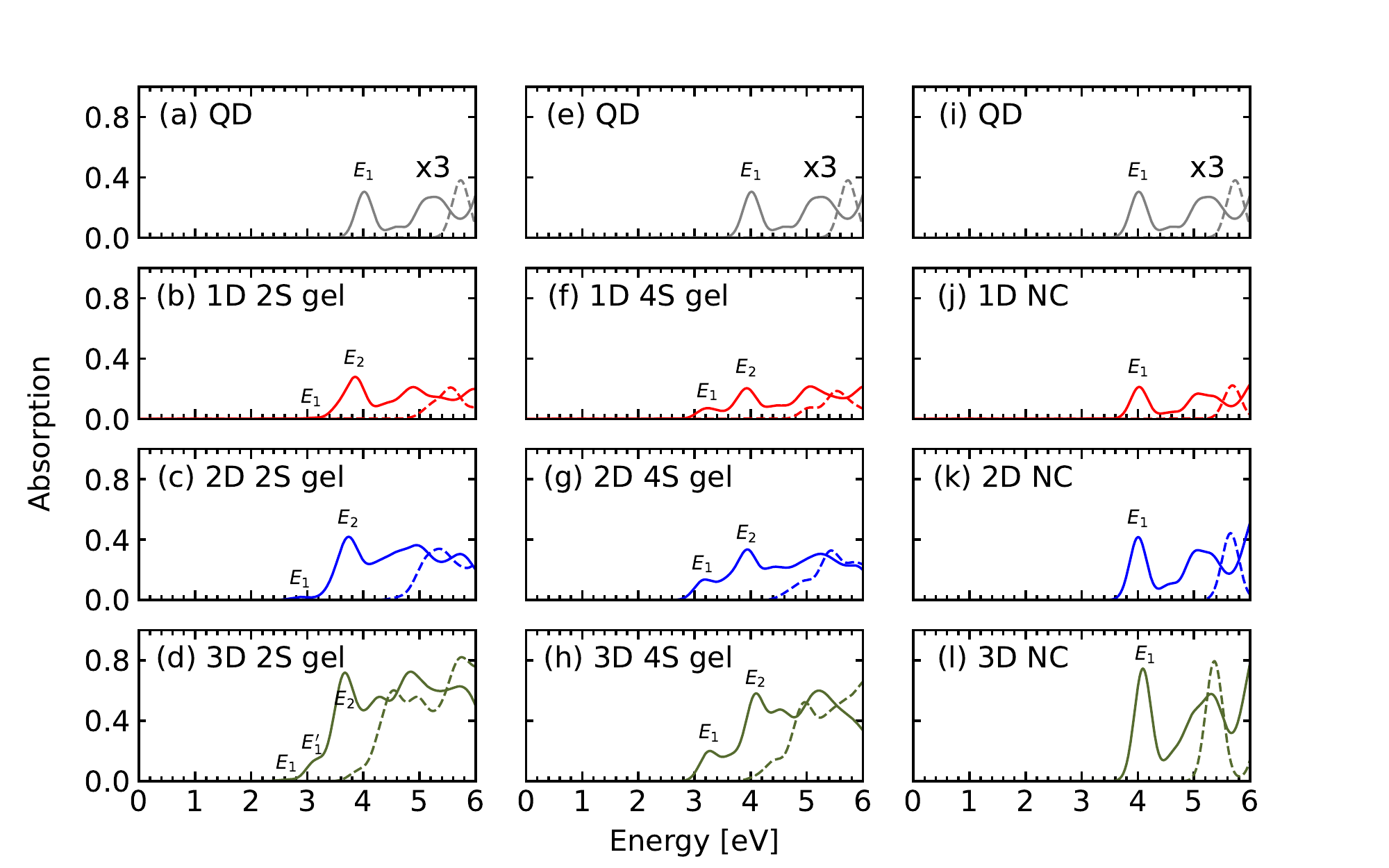}
\caption{Optical absorption spectra calculated from BSE (including electron-hole interactions, solid lines) and RPA (without electron-hole interactions, dashed lines) for (a,e,i) the discrete QD, (b-d) 2S gel in 1D, 2D, and 3D, respectively, (f-h) 4S gel in 1D, 2D, and 3D, respectively, and (j-l) NC in 1D, 2D, and 3D, respectively. We duplicate (a) as (e) and (i) to facilitate the comparison for each series across different dimensions. A 0.15 eV broadening is applied in all panels. The absorption intensities in (a,e,i) are magnified by three times.}
\label{fig:bse}
\end{figure*}

Once we understand the trends in the transport gap, we proceed with BSE calculations of the optical properties. Figure \ref{fig:bse} shows the absorption spectra calculated from BSE (including electron-hole interactions and capturing excitons, solid lines) and random-phase approximation (RPA, based on the $GW$ electronic structure and without electron-hole interactions, dashed lines). One can see that for every system, the lowest absorption peak calculated from BSE is well below the $GW$ band gap and the RPA absorption spectrum, indicating the formation of bound excitons. In this work, we focus on the major absorption peaks around and below 4 eV, which we mark as $E_1$ and $E_2$ in each panel of Figure \ref{fig:bse}. The only exception is 3D 2S gel, where we mark an additional $E_1'$ that has similar nature as $E_1$.

Table \ref{tab:gaps} lists the optical excitation energies for $E_1$ and $E_2$, as well as the exciton binding energies ($E_{\rm b}$) defined as the difference between $E_{\rm g}^{GW}$ and $E_1$. Our BSE result for the discrete QD agrees well with existing experiment: our calculations predict an optical gap of 4.02 eV, while using the empirical fitting formula from Ref. \citenum{YQGP03}, the optical gap of a 1.6 nm CdS QD is 3.85 eV. The difference might be attributed to the uncertainties in the measurement of the size, which is sensitive in determining the optical gap for ultra-small QDs.

Figure \ref{fig:bse} and Table \ref{tab:gaps} reveal trends in the optical properties. The most intense peak is $E_2$ for the gels and $E_1$ for the NCs. The $E_2$ of the 4S series and the $E_1$ of the NC series in all dimensions are similar (within 0.1 eV) in energy to the major absorption peak ($E_1$) of the discrete QD, albeit their quite different $GW$ band gaps. In the 2S series, a small red shift in energy is observed, where the change in $E_2$ as a function of the dimensionality is much smaller than the change in $E_{\rm g}^{GW}$. We note that the small red shift in the optical excitation energies compared to discrete QDs is consistent with prior experimental observations.\cite{DWE02, WTLH09} In addition to the major peak $E_2$, both the 2S and 4S gel series feature a satellite peak, which we denote by $E_1$ in Figure \ref{fig:bse}. $E_1$ is about 1 eV lower in energy than $E_2$ and follows the same trend as $E_2$ when the dimensionality increases. 

This weak dependence of optical excitation energies on the dimensionality can be understood via the trends in both $E_{\rm g}^{GW}$ and $E_{\rm b}$. As the dimensionality increases, $E_{\rm g}^{GW}$ decreases due to enhanced one-body coupling and many-body dielectric screening thanks to the presence of neighboring QDs, as we discussed above. For the same reason, $E_{\rm b}$ also decreases, by a similar amount. As a consequence, the changes in both $E_{\rm g}^{GW}$ and $E_{\rm b}$ roughly cancel each other, resulting in an $E_1$ that is weakly dependent on the dimensionality. The same argument holds for $E_2$. This phenomenon has been well understood in the context of a somewhat related but different scenario: consider the comparison between a single molecule and the same molecule adsorbed on a surface, where the latter acts as a dielectric environment. The fundamental gap of the adsorbed molecule is reduced compared to the isolated molecule, while the optical gap stays roughly unchanged \cite{SAVL21, DT19, S13}.

To understand the nature of these peaks and differentiate $E_1$ and $E_2$ for the covalent gels, we analyze the excited-state wavefunctions. In BSE, the excited-state wavefunction is a linear combination of transitions between Kohn-Sham orbitals, i.e., $\Psi\left(\mathbf{r}_{\rm e},\mathbf{r}_{\rm h}\right)=\sum_{vc}A_{vc}\phi_v^*(\mathbf{r}_{\rm h})\phi_c(\mathbf{r}_{\rm e})$. Here, $\Psi\left(\mathbf{r}_{\rm e},\mathbf{r}_{\rm h}\right)$ is the excited-state wavefunction, with $\mathbf{r}_{\rm e}$ ($\mathbf{r}_{\rm h}$) the position of the electron (hole). $\phi_v$ ($\phi_c$) is a valance (conduction) orbital from Kohn-Sham DFT, with $A_{vc}$ being the expansion coefficient for a specific $v\to c$ transition. For conciseness, we have omitted the $\mathbf{k}$-index in $A$, $\phi_v$, and $\phi_c$. $\Psi\left(\mathbf{r}_{\rm e},\mathbf{r}_{\rm h}\right)$ is a six-dimensional quantity, so we fix the $\mathbf{r}_{\rm h}$ to a point of our choice $\mathbf{R}_{\rm h}$ (see below) and plot the isosurface of the three-dimensional quantity $\Psi\left(\mathbf{r}_{\rm e};\mathbf{r}_{\rm h}=\mathbf{R}_{\rm h}\right)$.

%put the exciton plot here
\begin{figure}[htb]
\centering
\includegraphics[width=4in]{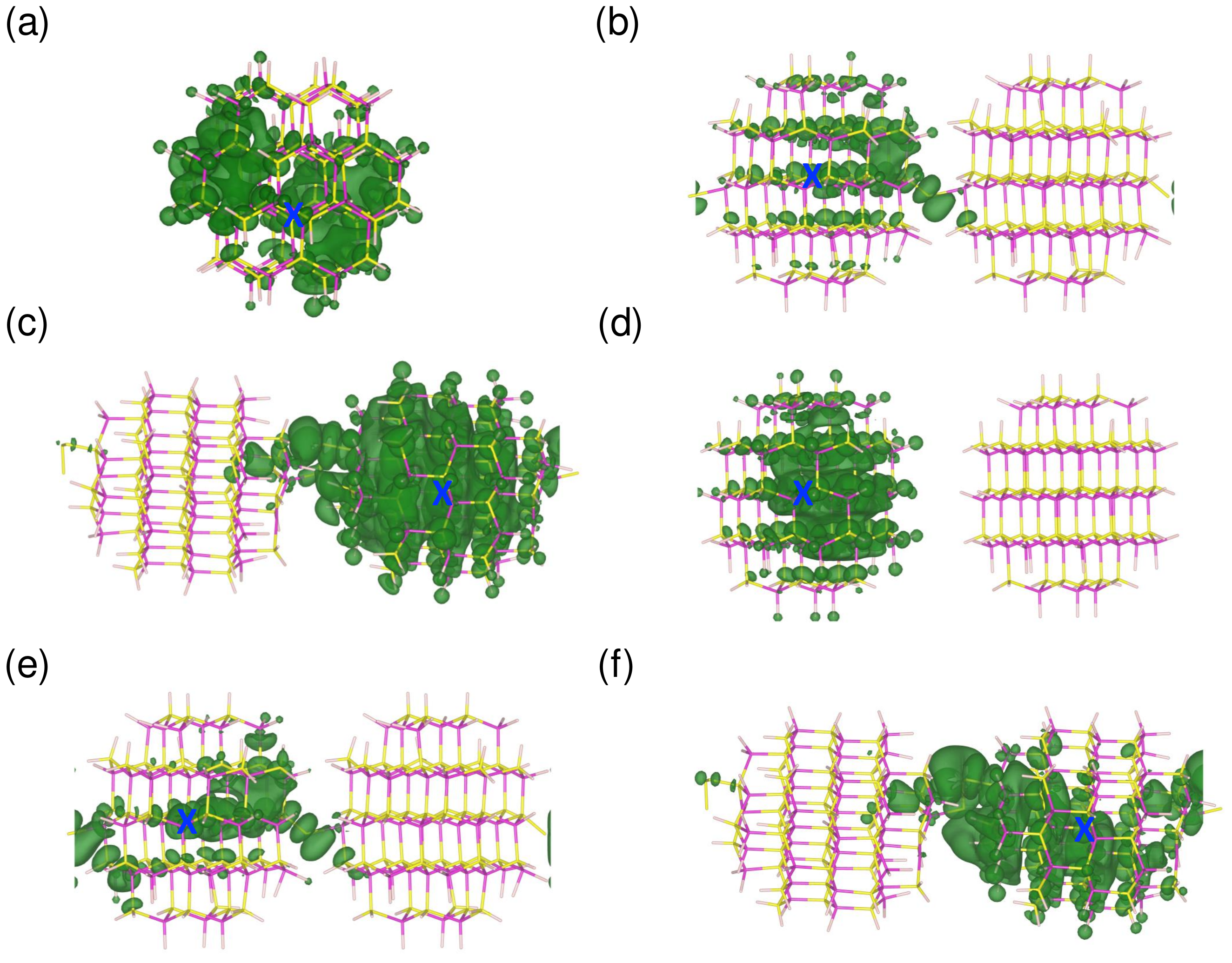}
\caption{Excited-state wavefunction $\Psi\left(\mathbf{r}_{\rm e};\mathbf{r}_{\rm h}=\mathbf{R}_{\rm h}\right)$ isosurface plots for (a) $E_1$ of the discrete QD; (b) $E_2$ of 1D 2S gel; (c) $E_2$ of 1D 4S gel; (d) $E_1$ of the 1D NC; (e) $E_1$ of 1D 2S gel; (f) $E_1$ of 1D 4S gel. In all panels, the blue ``X'' indicates the hole position, $\mathbf{R}_{\rm h}$.}
\label{fig:exciton}
\end{figure}

Figure \ref{fig:exciton} compares the excited-state wavefunctions for $E_1$ of the discrete QD (Figure \ref{fig:exciton}a), $E_2$ of 1D 2S gel (Figure \ref{fig:exciton}b), $E_2$ of 1D 4S gel (Figure \ref{fig:exciton}c), $E_1$ of the 1D NC (Figure \ref{fig:exciton}d), $E_1$ of 1D 2S gel (Figure \ref{fig:exciton}e), and $E_1$ of 1D 4S gel (Figure \ref{fig:exciton}f). In these plots, we adopt the same isosurface value for (b)-(d), and the same value for (e) and (f). The blue ``X'' indicates the hole position, $\mathbf{R}_{\rm h}$, and we have placed $\mathbf{R}_{\rm h}$ at the same place (near a sulfur atom in the center of the QD) for all systems to enable a better comparison. Figure S6 and Figure S7 show similar plots for the 2D and 3D gel structures, respectively.

From Figure \ref{fig:exciton}, one can see that the $E_2$ peaks of both 1D 2S and 1D 4S QD gels (Figure \ref{fig:exciton}b,c) can be assigned as ``bulk'' QD transitions, i.e., the excited-state wavefunction is largely localized on the QD (similar to $E_1$ of the discrete QD, Figure \ref{fig:exciton}a), with additional small contributions from the sulfur linker. The $E_1$ peak of 1D NC (Figure \ref{fig:exciton}d) resembles that of the discrete QD as well. This picture is in sharp contrast with the satellite peaks ($E_1$) in the covalent gels. Figure \ref{fig:exciton}e,f illustrate the nature of the satellite $E_1$ peak in the 1D 2S and 1D 4S gels: even when we place the hole in the center of the QD structure, the electron distribution still has a large weight near the covalent sulfur linker, with additional contributions from the QD. The presence of the satellite peak underscores the effects of the covalent linkers, which provide additional sites for excited-state formation and facilitate charge transfer between different QDs in the gel. We note that the $E_1$ absorption peak of the 1D 4S gel is more pronounced than that of the 1D 2S gel (Figure \ref{fig:bse}), consistent with the enhanced electronic distribution (Figure \ref{fig:exciton}). This is perhaps due to the larger number of unpassivated and zero-valance sulfur atoms in the 1D 4S gel, which produce more linker states within the gap of the otherwise pristine QD.

In summary, we have systematically compared the electronic and optical properties of two types of QD assemblies, the covalently bound QD gels and the van der Waals bound QD NCs, using first-principles $GW$-BSE approach within the framework of many-body perturbation theory. We showed how the properties evolve from those of a discrete QD, as the dimensionality of the assembly increases. We found that despite of the reduction in the quasiparticle band gap due to the many-body dielectric screening, the optical excitation energies corresponding to  QD-localized transitions stay roughly unchanged compared to a discrete QD. Moreover, we found that the covalently bound QD gels feature additional lower-energy peaks in the absorption spectra that can be assigned as transitions largely localized on the di- or tetra-sulfide linker groups, which are more prominent in the 4S gels than in the 2S gels. Physically, we have attributed these transitions to the presence of unpassivated, zero-valance sulfur atoms in the linker that give rise to ``mid-gap'' states. Our results provide a microscopic understanding of the electronic and optical properties of QD assemblies and unveil the difference between the two distinct types of assemblies, QD gels and QD NCs. We hope our work could shine light on the understanding of charge and energy transfer mechanisms in QD assemblies and future development of such materials.

\section{Supporting Information Description}
Computational parameters for all systems; Density of states for the 4S gel series and the NC series; DFT-LDA band structures for all systems; $GW$-BSE convergence studies; Excited-state wavefunction isosurface plots for the 2D gels and 3D gels.

\section*{Acknowledgements}
We thank Stephanie Brock, Long Luo, and Jier Huang for fruitful discussions. Z.-F.L. acknowledges an NSF CAREER Award, DMR-2044552. J.F. acknowledges a Rumble Fellowship and the A. Paul and Carole C. Schaap Endowed Distinguished Graduate Award in Chemistry at Wayne State University. DFT calculations in this work use computational resources at the Center for Nanoscale Materials, a U.S. DOE Office of Science User Facility, supported by the U.S. DOE, Office of Basic Energy Sciences, under Contract No. DE-AC02-06CH11357. Large-scale $GW$-BSE calculations are performed using resources of the National Energy Research Scientific Computing Center (NERSC), a U.S. Department of Energy Office of Science User Facility located at Lawrence Berkeley National Laboratory, operated under Contract No. DE-AC02-05CH11231 through NERSC award BES-ERCAP0020328, as well as the Extreme Science and Engineering Discovery Environment (XSEDE), which is supported by NSF grant number ACI-1548562 through allocation PHY220043.

\bibliography{refs.bib}

%\begin{tocentry}
%\center
%\includegraphics{TOC.pdf}
%\end{tocentry}

\end{document}